\documentclass[sort&compress]{elsarticle}

\usepackage[margin=1in]{geometry}
\usepackage{graphicx}
\usepackage[american]{babel}
\usepackage[utf8]{inputenc}
\usepackage[T1]{fontenc}
\usepackage{latexsym,amsmath,amssymb}
\usepackage{booktabs}
\usepackage{float}
\usepackage{url}
\allowdisplaybreaks
\usepackage[sc]{mathpazo}
\usepackage{color}
\usepackage{caption}

\usepackage{color}

\usepackage{cleveref}

\usepackage{nomencl}
\usepackage{framed} 
\makenomenclature

\begin{document}

\begin{frontmatter}
	
\title{Cost-optimal design of a simplified highly renewable \\	Chinese electricity network}

\author[label2]{Hailiang Liu}
\ead{liuhailiang1989@gmail.com}
\author[label2]{Gorm Bruun Andresen}
\author[label2]{Martin Greiner}
\address[label2]{Department of Engineering, Aarhus University, Inge Lehmanns Gade 10, 8000 Aarhus C,  Denmark}

\begin{abstract}

Rapid economic growth in China has lead to an increasing energy demand in the
country. In combination with China's emission control and clean air initiatives, it has
resulted in large-scale expansion of the leading renewable energy technologies, wind
and solar power. Their intermittent nature and uneven geographic distribution,
however, raises the question of how to best exploit them in a future sustainable
electricity system, where their combined production may very well exceed that of all
other technologies. It is well known that interconnecting distant regions provides more
favorable production patterns from wind and solar. On the other hand, long-distance
connections challenge traditional local energy autonomy. In this paper, the advantage
of interconnecting the contiguous provinces of China is quantified. To this end, two
different methodologies are introduced. The first aims at gradually increasing
heterogeneity, that is non-local wind and solar power production, to minimize
production costs without regard to the match between production and demand. The
second method optimizes the trade-off between low cost production and high utility
value of the energy. In both cases, the study of a 100\% renewable Chinese electricity
network is based on 8 years of high-resolution hourly time series of wind and solar
power generation and electricity demand for each of the provinces. From the study we
conclude that compared to a baseline design of homogeneously distributed renewable capacities, a heterogeneous network not only lowers capital investments but also reduces backup
dispatches from thermal units. Installing more capacity in provinces like
Inner Mongolia, Jiangsu, Hainan and north-western regions, heterogeneous layouts
may lower the levelized cost of electricity (LCOE) by up to 27\%, and reduce backup
needs by up to 64\%.

\end{abstract}


\end{frontmatter}

\newpage

\begin{framed}
\nomenclature{$LCOE$}{Levelized cost of electricity}
\nomenclature{$L_n$}{Load}
\nomenclature{$ G_n^R $}{Renewable power generation}
\nomenclature{$ G_n^W $}{Wind power generation}
\nomenclature{$ G_n^S $}{Solar power generation}
\nomenclature{$ \gamma_n $}{Renewable penetration level}
\nomenclature{$ \alpha_n $}{Share of average wind power generation in the wind-solar mix}
\nomenclature{$ \Delta_n $}{Mismatch between renewable generation and load}
\nomenclature{$ P_n $}{Injection pattern}
\nomenclature{$ C_n $}{Curtailment}
\nomenclature{$ G_n^B $}{Backup power generation}
\nomenclature{$ B_n $}{Balancing}
\nomenclature{$ F_l $}{Power flow in link $ l $}
\nomenclature{$ H_{ln} $}{Matrix of power transfer distribution factors between node $ n $ and link $ l $}
\nomenclature{$ \mathcal{K}_n^W $}{Wind power capacity}
\nomenclature{$ \mathcal{K}_n^S $}{Solar power capacity}
\nomenclature{$ CF_n^{W} $}{Wind power capacity factor}
\nomenclature{$ CF_n^{S} $}{Solar power capacity factor}
\nomenclature{$ E^B $}{Backup energy}
\nomenclature{$ \mathcal{K}^B $}{Backup capacity}
\nomenclature{$ \mathcal{K}^T $}{Transmission capacity}
\nomenclature{$ d_l $}{Length of link $ l $}
\nomenclature{$ OpEx $}{Operation expenditure}
\nomenclature{$ CapEx $}{Capital expenditure}
\nomenclature{$ r $}{rate of return}
\nomenclature{$ K $}{Heterogeniety parameter in optimized layouts}
\nomenclature{$ \beta $}{Heterogeniety parameter in heuristic layouts}
\nomenclature{$ n $}{A province}
\printnomenclature
\end{framed}

\section{Introduction}

China is undergoing tremendous challenges of decarbonization and  air quality impairment due to rapid industrial growth, transportation expansion and sharply increased demands of electricity  \cite{huang2012typical}. Coal as the main source of power is being changed, and the employment of alternative sources has become an important part of the Chinese energy policy. For many years, the only renewable energy source in China has been hydro, meeting 19\% of annual electricity demand in 2014 \cite{yearbook}. But hydro is reaching its full potential due to site limitations. Wind and solar power, however, have become affordable  \cite{Analysis} and more suitable for large scale expansion  \cite{brown,he2014and,he2016and}.

Considering the growing penetration of renewables, some scholars have looked into implications of integrating large amounts of wind and solar energy in the Chinese power sector. A combined source-grid-load planning model \cite{zhang2017integrated} was introduced to seek a cost-optimal solution at the macro level, taking into account higher renewable penetration up to 2030. They suggested striking a balance between resource-rich and high-load regions by means of a rapid expansion of the inter-regional transmission grid. 
The same is suggested by \cite{yi2016inter}. In this context, economic savings and better utilization of wind and solar power may be achieved by shifting renewable capacities towards resource-rich regions that are linked to regions with high demands by high-capacity transmission lines. In the studies, emphasis was given to policy target driven scenarios and system cost reductions were not explicitly discussed.

Both \cite{zhang2017integrated} and \cite{yi2016inter} studied the renewable integration using annual average values. This does not explicitly capture the variable nature of renewables. Ref. \cite{huber2015optimal} is the first study to base the analysis on hourly data. The high temporal resolution allowed insights into backup, storage and flexibility needs. All three studies were concerned with China as a single aggregated entity, rather than a power system with inter-connecting provinces.

\begin{figure}[h!]
	\centering
	\includegraphics[width=0.9\linewidth]{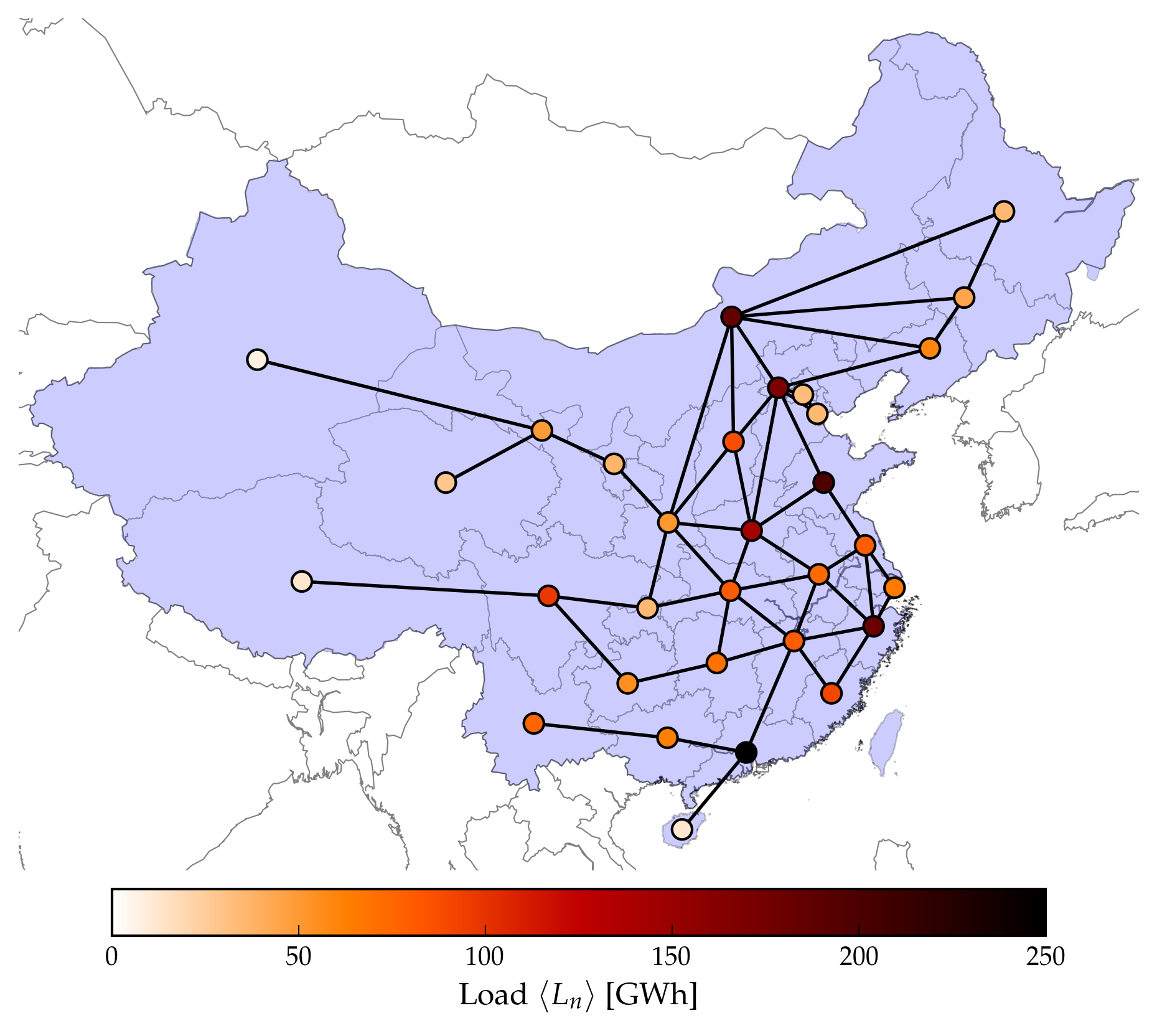}
	\caption{The 31 contiguous provinces of China connected in a future Chinese electricity network, simplified from the TSO's plan \cite{wiki:network}. A colored disc in each region indicates the average hourly electricity demand in 2050, which is predicted based on GDP per capita forecasts. Details are presented in the supplementary material.}
	\label{fig:load}
\end{figure}

This paper is inspired by similar studies for large scale integration of wind and solar power in Europe \cite{rodriguez2015cost,schlachtberger2017benefits}, US \cite{becker2014features,becker2015renewable} and Australia \cite{prasad2017assessment}, where the analysis are implemented on a continental level, yet still with starting points of high spatial and temporal resolution wind-solar generation time series. As an emerging titan, the Chinese power sector is growing more and more renewable, but for lack of high resolution data, studies for this new regime has yet to appear. The present paper focuses on a simplified 100\% renewable power system for the 31 contiguous provinces of China. Hourly time series of wind and solar power generation, as well as load covering eight years for each of the 31 provinces are generated and used in the analysis of the interconnected network, the structure of which is illustrated in Figure \ref{fig:load}. The time series are based on high quality weather data and have been validated to the extend that it was possible using state-of-the-art practices \cite{andresen2015validation,staffell2016using,pfenninger2016long}. In our opinion, our primary addition to the literature is filling the gap with this unique validated high resolution dataset in the new domain and laying the ground work for other interested researchers.

The first part of the analysis consists of two ambitious baseline scenarios for the far future, in 2050, where the overall cost composition of the Chinese power system with a 100\% wind-solar penetration is analyzed without and with an interconnecting transmission grid between the provinces. In the baseline scenarios all provinces have similar energy autonomy in the sense that the wind-solar penetration is fixed to 100\% for each individual province. 

To analyze the economical advantage of relocating wind and solar generators to more favorable locations, two additional scenarios are analyzed. The first of these is based on a heuristic, analogous to \cite{thesis:leon},  where sites with better capacity factors, i.e. higher annual wind or solar yield, gradually produce a higher fraction of the total energy. However, this approach ignores the match between demand and generation, and may lead to relative high curtailment and balancing energy needs. This is taken into account in the second approach, where the total cost per MWh energy delivered to cover the demand is minimized by redistributing wind and solar capacity to favorable locations. The second approach is a trade-off between lowering the wind and solar energy production cost and maximizing the useful energy from these variable resources. It is achieved by selecting locations with high energy yields as well as production patterns that complement each other to provide a better match to the demand time series. Again, the method allows different degrees of heterogeneity as decision makers may want to factor in energy autonomy as well as cost optimization. The difference between the two approaches highlights the importance of taking into account correlations in highly renewable scenarios. Table \ref{table:scenerios} summaries the four scenarios considered throughout the paper, highlighting their differences.

\begin{table}[h!]
	\centering
	\caption{List of scenarios considered in this study. Two homogeneous designs as the baselines without or with transmission, and two heterogeneous ones based on  heuristic and optimized layouts. The layouts are characterized by wind and solar power mix $ \alpha_n $ of Equation \ref{eq:alpha} and renewable penetration $ \gamma_n $ of Equation \ref{eq:gamma} for provinces $ n $}
	\label{table:scenerios}
	\begingroup
	\renewcommand{\arraystretch}{1.4} 
	\begin{tabular}{@{}lllll@{}}
		\toprule
		& \multicolumn{2}{c}{Homogeneous designs}                            & \multicolumn{2}{c}{Heterogeneous designs}    \\ 
		\cmidrule(lr){2-3} \cmidrule(lr){4-5} 
		\textbf{} & \multicolumn{1}{l}{No transmission} & \multicolumn{1}{l}{With transmission} & \multicolumn{1}{l}{Heuristic layouts}  & \multicolumn{1}{l}{Optimized layouts}  \\ 
		\midrule 
		$\alpha_n $ & $ [0, 1] $, same $ \forall n $ & $ [0, 1] $, same $ \forall n $ & $ [0, 1]$, varies among $n$        & $ [0, 1] $, varies among $n$                   \\
		$ \gamma_n $ & $ 1,  \forall n $               & $ 1, \forall n $                & $ [0, +\infty] $, depends on $ \beta$ & $ \frac{1}{K} \leqslant \gamma_n \leqslant K $                                               
		\\ \bottomrule
	\end{tabular}
	\endgroup
\end{table}

This paper is organized as follows. Section \ref{methods} introduces hourly wind, solar and load time series as well as methods of quantifying backup and inter-regional transmission needs. Section \ref{homo} presents the homogeneous baseline scenarios. In Section \ref{hetero} wind and solar generators are relocated to more favorable provinces using the two approaches to lower total system cost including backup units and transmission grid. Sensitivity analysis and the issue of curtailment is discussed in Section \ref{disc} before Section \ref{conc} concludes the paper.

\section{Data and methods}
\label{methods}

A simplified model of the Chinese electricity network is used for the study. Here, each province is aggregated into a single node located at its geometric center, and the connecting links represent the combined transmission capacity between neighboring provinces. Figure \ref{fig:load} shows the network topology and the average loads of the individual provinces. 

\begin{figure}[h!]
	\centering
	\includegraphics[width=0.95\linewidth]{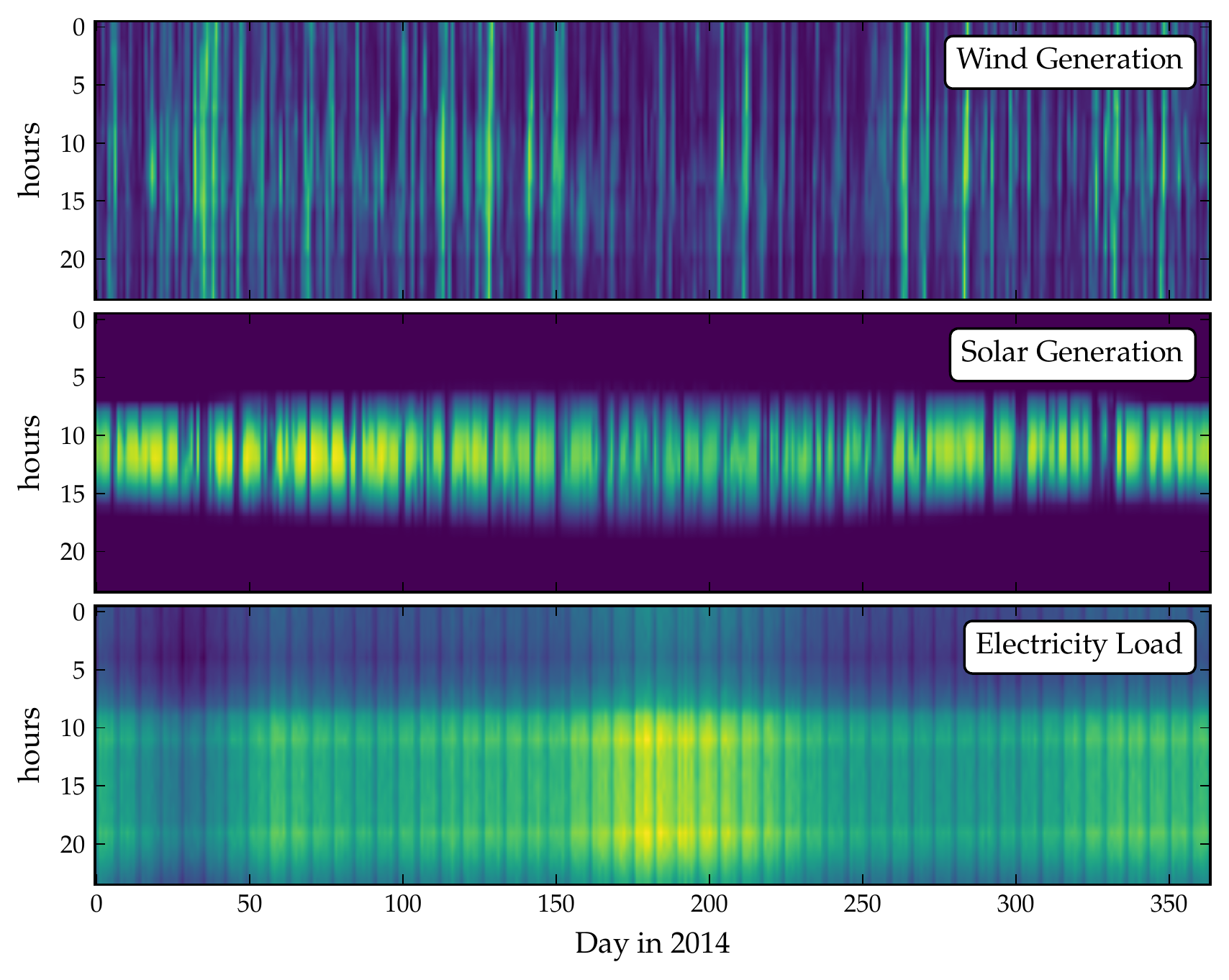}
	\caption{Temporal patterns of wind (top), solar power generation (middle) and electricity demand (bottom) for the province Jiangsu in the year of 2014. The high resolution profile is modeled by combining historical provincial installed renewable capacities in 2014 \cite{wind2015} and hourly weather data in the Renewable Energy Atlas \cite{andresen2015validation}.}
	\label{fig:temporal_patterns}
\end{figure}

The hourly renewable power generation in node $ n $ is composed of wind $G_n^W$ and solar PV  $G_n^S$ generation:
\begin{equation}
G_n^R(t)=G_n^W(t)+G_n^S(t).
\end{equation}
This renewable generation, shown in Figure~\ref{fig:temporal_patterns}, is modeled using hourly weather data covering 2005--2012 with spatial resolution of $ 40 \times 40  $~km$^2$. Details can be found in the supplementary material. The hourly load $ L_n$ time series  modeling is also described there.

The penetration $ \gamma_n $ of renewable power generation for each node is defined as the ratio between average renewable generation, ignoring curtailment, and average load $L_n$:
\begin{equation}\label{eq:gamma}
\gamma_n = \frac{\langle G_n^R \rangle}{\langle L_n \rangle}.
\end{equation}
Furthermore, the wind-solar mix $0\leq\alpha_n\leq1$ is defined as
\begin{equation}\label{eq:alpha}
 \alpha_n = \frac{\langle G_n^W \rangle}{\langle G_n^R \rangle}.
\end{equation}
For example, $\alpha_n=0.6$ indicates that 60\% of the average renewable generation originates from wind and 40\% from solar ($ 1-\alpha_n $).

In general, a nodal mismatch between renewable generation and load will occur for most hours. This is calculated as
\begin{equation}
	\Delta_n(t)=G^R_n(t)-L_n(t)\,.
\end{equation}
The mismatch has to be balanced by power exports and imports $ P_n(t) $ among the nodes, dispatch of nodal backup generation $ G_n^B(t) $ or  curtailment $ C_n(t) $ of the wind and solar PV generators. This nodal balance can be expressed as
\begin{equation}
\label{eq:mismatch}
G_n^R(t) - L_n(t)=P_n(t)+C_n(t)-G_n^B(t),
\end{equation}
where the quantities on the left hand side are defined exogenously in the model while those on the right hand side are determined by means of maximizing the utilization of renewable energy given the system constraints as described in \cite{rodriguez2014transmission}. The combined backup dispatch and curtailment is called balancing. It is defined as
\begin{equation}
B_n(t)=C_n(t)-G_n^B(t)~,
\end{equation}
and can take both positive and negative values.  By definition, $ G_n^B(t) $ and $ C_n(t) $ are both non-negative for each time step. Please note, although backup energy from conventional sources are needed in the network, we still call the system 100\% powered by wind and solar energy. This is because the renewable penetration level 100\% is defined as the ratio of average renewable generation to the average load. Conventional power is needed at times where wind and solar power can not meet the power demand, due to their intermittent nature.

For simplicity we follow the approach of synchronized balancing \cite{rodriguez2015cost}, where the global mismatch is distributed to the individual nodes proportional to their average load:
\begin{equation}
\label{eq:sync}
B_{n}(t) = \left[\sum_{m} \Delta_{m}(t)\right]\frac{\langle L_n \rangle}{\sum_{k} \langle L_{k}\rangle}~.
\end{equation}
This choice of a balancing scheme fixes the exports and imports at the individual nodes as
\begin{equation}
P_{n} = \Delta_{n}(t) - \frac{\langle L_n \rangle}{\sum_{k} \langle L_{k}\rangle} \sum_{m}\Delta_{m}(t)~.
\end{equation}
The resulting power flows $ F_l $ on the links $ l $ can be expressed using the matrix of power transfer distribution factors $ H_{ln} $ (PTDF matrix) \cite{thesis:leon}:
\begin{equation}
F_{l} = \sum_{n}H_{ln}P_{n}~,
\end{equation}
where we use the index $ l=l(m, n) $ for the link between nodes $ m $ and $ n $.

\subsection{Renewable Capacities}

The installed capacities for wind $ \mathcal{K}^W_n $ and solar $ \mathcal{K}^S_n $ are derived from the parameters $ \gamma_n $ and $ \alpha_n $ as follows
\begin{equation}
\mathcal{K}_n^W=\dfrac{\gamma_n\alpha_n\langle L_n \rangle}{CF_n^W},
\end{equation}
\begin{equation}
\mathcal{K}_n^S=\dfrac{\gamma_n(1-\alpha_n)\langle L_n \rangle}{CF_n^S},
\end{equation}
where $ CF_n^W $ and $ CF_n^S $ denote the nodal wind and solar capacity factors. These are shown in Figure \ref{fig:CF_on_map}, listed in Table \ref{tableCF} and discussed further in the supplementary material.

\begin{figure*}[h!]
	\centering
	\includegraphics[width=0.495\linewidth]{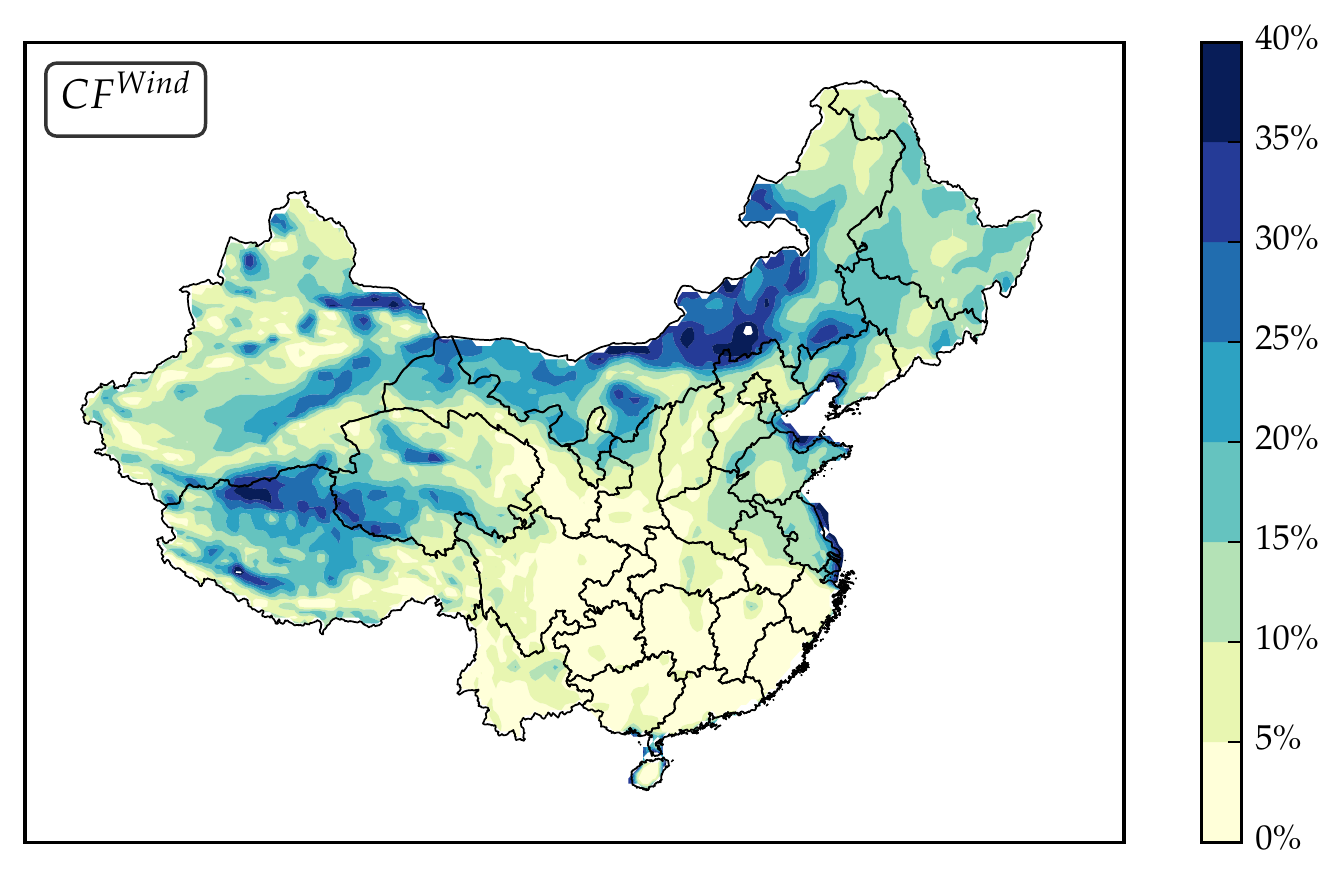}
	\includegraphics[width=0.495\linewidth]{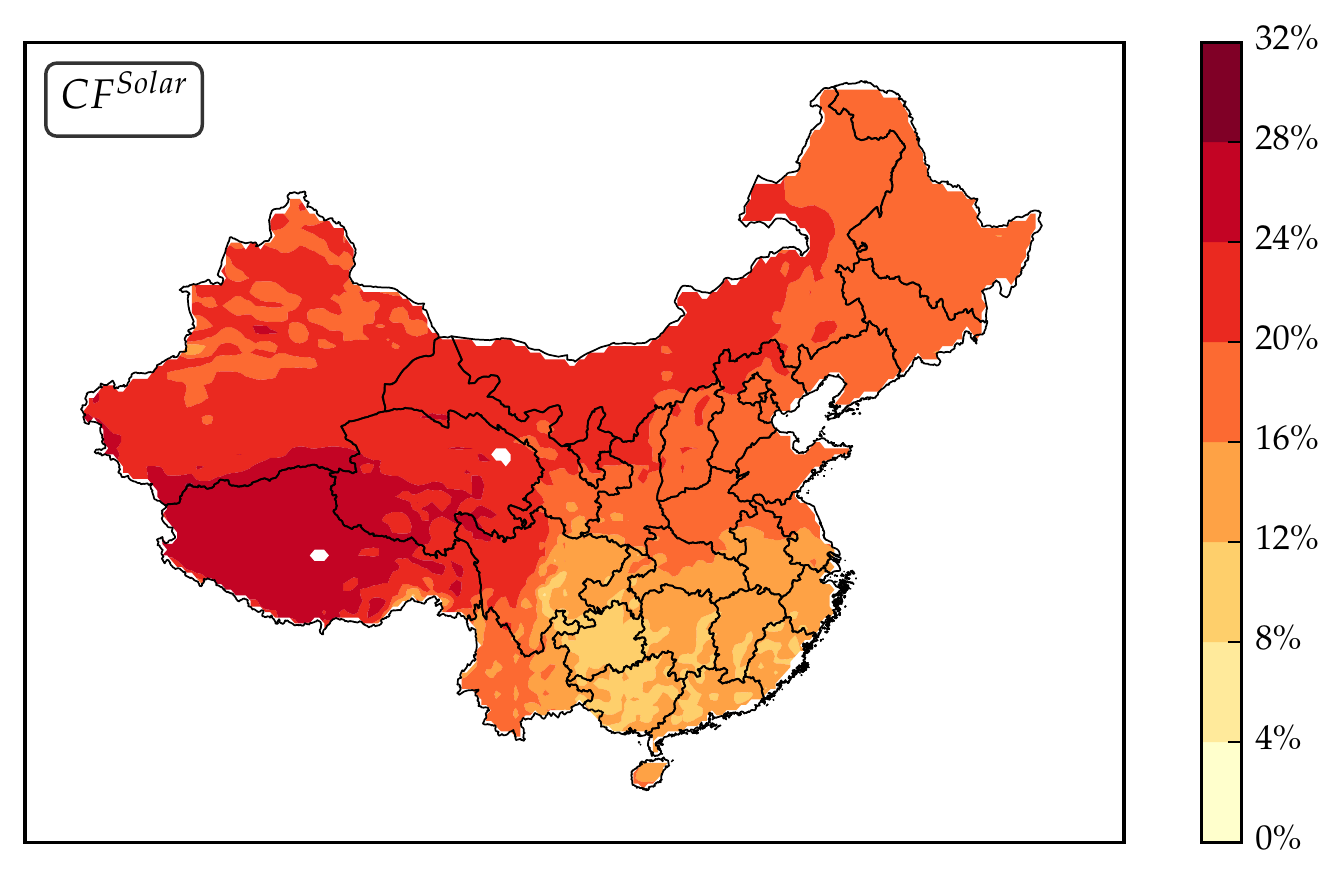}
	\caption{Maps showing the capacity factor $ CF $ for wind (left) and solar (right) throughout the 31 contiguous provinces of China. The calculations are based on Renewable Energy Atlas \cite{andresen2015validation} using the weather year of 2012. The results are in good agreement with the corresponding maps presented in Ref. \cite{he2014and,he2016and}.}
	\label{fig:CF_on_map}
\end{figure*}
 
\begin{table*}[h!]
	\centering	
	\caption{Wind and solar capacity factors, based on the Renewable Energy Atlas \cite{andresen2015validation}, and predicted 2050 hourly average load [GWh], which is calculated by combining load and GDP per capita relations \cite{lin2016economic} with OECD's economic forecast \cite{Domestic82:online}. More details are included in the supplementary material.}
	\begin{tabular}{@{}lrrr|lrrr@{}}
		\toprule
		\textbf{Provinces} & \multicolumn{1}{l}{$\mathbf{CF}^W_n$} & \multicolumn{1}{l}{$\mathbf{CF}^S_n$} & \multicolumn{1}{l|}{$\mathbf{\langle L_n \rangle}$} & \textbf{Provinces} & \multicolumn{1}{l}{$\mathbf{CF}^W_n$} & \multicolumn{1}{l}{$\mathbf{CF}^S_n$} & \multicolumn{1}{l}{$\mathbf{\langle L_n \rangle}$} \\ 
		\midrule
		Anhui & 0.176 & 0.160 & 72.6 & Jilin & 0.261 & 0.189 & 44.2 \\
		Beijing & 0.173 & 0.192 & 32.8 & Liaoning & 0.300 & 0.191 & 59.3\\
		Chongqing & 0.105 & 0.154 & 35.0 & Inner Mongolia & 0.351 & 0.206 & 184.3 \\
		Fujian & 0.161 & 0.135 & 89.6 &  Ningxia & 0.264 & 0.203 & 36.8 \\
		Gansu & 0.250 & 0.213 & 48.8 & Qinghai & 0.289 & 0.246 & 28.2 \\
		Guangdong & 0.191 & 0.142 & 248.6 & Shaanxi & 0.216 & 0.193 & 50.9\\
		Guangxi & 0.171 & 0.136 & 62.5 & Shandong & 0.292 & 0.181 & 198.7 \\
		Guizhou & 0.138 & 0.126 & 54.9 & Shanghai & 0.457 & 0.152 & 64.3 \\
		Hainan & 0.312 & 0.157 & 12.5 & Shanxi & 0.197 & 0.191 & 87.5 \\
		Hebei & 0.290 & 0.194 & 167.4 & Sichuan & 0.231 & 0.218 & 97.0 \\
		Heilongjiang & 0.240 & 0.183 & 34.9 & Tianjin & 0.230 & 0.186 & 34.9 \\
		Henan & 0.181 & 0.177 & 142.8 & Xinjiang & 0.247 & 0.224 & 7.3 \\
		Hubei & 0.156 & 0.163 & 80.6 & Tibet & 0.300 & 0.268 & 12.2\\
		Hunan & 0.142 & 0.137 & 69.8 & Yunnan & 0.160 & 0.158 & 75.2 \\
		Jiangsu & 0.329 & 0.167 & 79.5 & Zhejiang & 0.171 & 0.140 & 181.2 \\
		Jiangxi & 0.131 & 0.138 & 80.6 &  & \multicolumn{1}{l}{} & \multicolumn{1}{l}{} & \multicolumn{1}{l}{} \\ 
		\bottomrule
	\end{tabular}
	\label{tableCF}
\end{table*}

\subsection{Heuristic layouts}

The heuristic layout assigns resources proportional to the CF, or more general to the CF raised to a power $ \beta $,
\begin{equation}\label{eq:betalayout1}
\gamma_{n}^{W} = \dfrac{(CF_{n}^{W})^{\beta}\langle L_{CN} \rangle}{\sum_m (CF_{m}^{W})^{\beta}\langle L_m \rangle}\gamma_{CN}, 
\end{equation}
where $ \gamma_{CN}=\sum_{n}\frac{\langle L_n \rangle}{\langle L_{CN} \rangle}\gamma_n $ is the overall penetration and $ \langle L_{CN} \rangle = \sum_{n}^{} \langle L_n \rangle $ is the overall average load of China, $ \gamma_{CN} $ is constrained to 1. This gives the renewable penetration for a wind-only layout. Analogously, the solar-only layout is,
\begin{equation}\label{eq:betalayout2}
\gamma_{n}^{S} = \dfrac{(CF_{n}^{S})^{\beta}\langle L_{CN} \rangle}{\sum_m (CF_{m}^{S})^{\beta}\langle L_m \rangle}\gamma_{CN}.
\end{equation}
The so-called $ \beta $-layout \cite{thesis:leon} can be constructed as a linear combination of the wind-only and solar-only layouts by setting,
\begin{equation}\label{eq:betalayout3}
\gamma_n = \alpha_{CN} \gamma_n^W + (1-\alpha_{CN})\gamma_n^S, 
\end{equation} 
and
\begin{equation}\label{eq:betalayout4}
\alpha_n = \alpha_{CN} \frac{\gamma_n^W}{\gamma_n}.
\end{equation}
Note that $\beta$ enters the two equations above via the definitions of $\gamma^W_n$ and $\gamma^S_n$. $ \beta=0 $ defines the homogeneous baseline design, while $ \beta=1, 2, 3 $ allow provinces with higher capacity factors to have bigger penetration levels $ \gamma_n $ than  provinces with smaller capacity factors.

\subsection{Optimized layouts}

In order to introduce an optimal heterogeneity in a way that includes the trade-off between favorable production patterns, capacity factors and low transmission grid costs, location of wind and solar generators, i.e., $ \alpha_n $ and $ \gamma_n $, must be optimized with the objective to minimize the total LCOE. To solve this non-linear optimization problem, a numerical algorithm based on a  Greedy Axial Search (GAS) \cite{thesis:leon} was chosen. It can deal with $ 2N=62 $ variables: $ \gamma_1,..., \gamma_N $ and $ \alpha_1,...,\alpha_N $ for the $ N=31 $ provinces.
The optimization problem is formulated as follows,
\begin{equation}
\begin{aligned}
& \underset{\underset{(n = 1, 2,\ldots, 31)}{\gamma_n, \; \alpha_n}}{\text{minimize}}
& & LCOE_{system} \\
& \text{subject to}
& & 0 \leqslant \alpha_n \leqslant 1,\; (n = 1, 2,\ldots, 31), \\
&&& \frac{1}{K} \leqslant \gamma_n \leqslant K,\; (n = 1, 2,\ldots, 31), \\
&&& \gamma_{CN} = 1.
\end{aligned}
\end{equation}
In the optimization process, the heterogeneity parameter $ K $ acts as a bound, keeping penetration levels away from extreme values. Installing too many renewables in very few provinces could be politically unattractive. For instance, provinces with relatively poor weather resources may wish to maintain some degree of energy autonomy. $ K = 1 $ implies $ \gamma_n = 1.0 $ for all provinces, $ K = 2 $ keeps $ \gamma_n $ within the range of 0.5 -- 2.0.

\subsection{Infrastructure measures}

Measures of the infrastructure are defined as follows. Backup energy $ E^B $ is defined as the nodal sum of the time averages of the backup generation
\begin{equation}
E^B=\sum_{n}^{}\langle G_n^B \rangle,
\end{equation}
and is interpreted as the total generation from conventional resources like hydro, nuclear, gas or coal. For simplicity, it is represented only by the popular Combined Cycle Gas Turbines (CCGT). The nodal backup power capacity $ \mathcal{K}_n^B $ is defined as a high quantile
\begin{equation}
q_n=\int_{0}^{\mathcal{K}_n^B}\mathrm{d}G_n^B p_n(G_n^B),
\end{equation}
of the backup generation distribution $p_n(G_n^B)$.  As in \cite{rodriguez2015localized,tranberg2015power}, the 99\% quantile is chosen for $ q_n $. Analogously, the transmission capacity $ \mathcal{K}_l^T $ of link $ l $ is expressed as
\begin{equation}
q_l=\int_{0}^{\mathcal{K}_l^T}\mathrm{d}|F_l|\,p_l(|F_l|),
\end{equation}
where $|F_l|$ denotes the absolute value of the flow on link $ l $. Again the 99\% quantile is chosen.

The total backup power capacity of the system is represented by
\begin{equation}
\mathcal{K}^B=\sum_{n}^{}\mathcal{K}_n^B,
\end{equation}
and the total transmission capacity is calculated as
\begin{equation}
\mathcal{K}^T=\sum_{l}^{}\mathcal{K}_l^Td_l,
\end{equation}
where $ d_l $, listed in the supplementary material, is the length of link~ $ l $.

\subsection{Objective function}

Levelized cost of electricity $ LCOE $ is the total cost per consumed energy unit calculated over the full lifetime of the infrastructure assets. It includes capital costs $ CapEx $ as well as operational expenses $ OpEx $ for the following investments: onshore and offshore wind turbines, solar PV, transmission and backup units. Backup expense is broken down into backup capacity and backup energy, and its cost is represented by that of a Combined Cycle Gas Turbines (CCGT). See Table \ref{table:costass}. 

\begin{table}[h!]
	\caption{Cost assumptions for different assets are separated into capital expenditure $ CapEx $ and operational expenditures $ OpEx $. The latter is divided into a fixed component that is independent of use and a variable component that depends on usage of the asset, e.g. fuel consumption. Data from  \cite{Analysis,thesis:leon}, in good agreement with  \cite{ouyang2014levelized}.}
	\label{table:costass}
	\setlength\tabcolsep{-2pt} 
	\begin{tabular*}{\columnwidth}{@{\extracolsep{\fill}}lrrrr@{}}
		\toprule
		\textbf{Asset} & \multicolumn{1}{r}{\begin{tabular}[r]{@{}r@{}}{CapEx}\\ \footnotesize{[}CNY(EUR)/W{]}\end{tabular}} & \multicolumn{1}{l}{\begin{tabular}[c]{@{}c@{}}${\mathrm{OpEx}_{Fix}}$\\ \footnotesize{[}CNY(EUR)/kW/y{]}\end{tabular}} & \multicolumn{1}{l}{\begin{tabular}[c]{@{}c@{}}${\mathrm{OpEx}_{var}}$\\ \footnotesize{[}CNY(EUR)/MWh{]}\end{tabular}} & \multicolumn{1}{l}{\begin{tabular}[l]{@{}r@{}}{Life time}\\ \footnotesize{[}years{]}\end{tabular}} \\
		\midrule
		{CCGT backup} & 6.6 (0.88) & 33 (4.40) & 410 (54.67) & 30 \\
		{Solar PV} & 8.2 (1.09) & 117 (15.60) & 0 & 25 \\
		{Onshore Wind} & 8.6 (1.15) & 220 (29.33) & 0 & 25 \\
		{Offshore Wind} & 14.7 (1.96) & 660 (88.00) & 0 & 30 \\ 
		\bottomrule
	\end{tabular*}
\end{table}

For any specific investment, we have
\begin{equation}
\mathrm{LCOE_V}=\dfrac{\mathrm{CapEx} + \sum_{t=1}^{lifetime}\dfrac{\mathrm{OpEx}_t}{(1+r)^t}}{\sum_{t=1}^{lifetime}\dfrac{L_t}{(1+r)^t}},
\end{equation}
where the time $ t $ is in units of year and runs until the lifetime of the asset $ V $, $ r $ is the rate of return, assumed to be 4\% per year, based on the current feed-in-tariff and its future predictions \cite{ouyang2014impacts,ouyang2014levelized}.
The levelized system cost, which is the objective function is defined as the sum over all assets:
\begin{equation}
\mathrm{LCOE_{system}}=\sum_{V}^{}\mathrm{LCOE_V},
\end{equation}

The levelized system cost of meeting the demand was calculated such that $L_t$ is equal to the electrical demand and the expenses are associated with the combined $ CapEx $ and $ OpEx $ for the backup generators, wind and solar generators and the transmission network. The values for the first three assets are provided in Table~\ref{table:costass}, while transmission costs are calculated as follows: Long-range high voltage AC lines costs 5000 CNY/km/MW, while DC lines have an average of 2000 CNY/km/MW on top of converter costs 1.125 million CNY \cite{wangweb}, and both have a life span of 40 years.

\section{Baseline design: The homogeneous case without / with transmission}
\label{homo}

The baseline design consists of a system with a homogeneous allocation of the wind and solar generators in the sense that all nodes are assigned a renewable penetration of 100\% ($ \gamma_n = 1\; \forall n$). This means that each of the provinces have installed wind and solar capacities that on average would be able to meet their load if the generation-load mismatch $\Delta_n$ could be ignored. The wind-solar mix is assumed to be identical for all provinces ($ \alpha_n = \alpha \;\forall n $), and the specific value of the mix is initially considered a free parameter.

\begin{figure}[h!]
	\centering
	\includegraphics[width=0.6\linewidth]{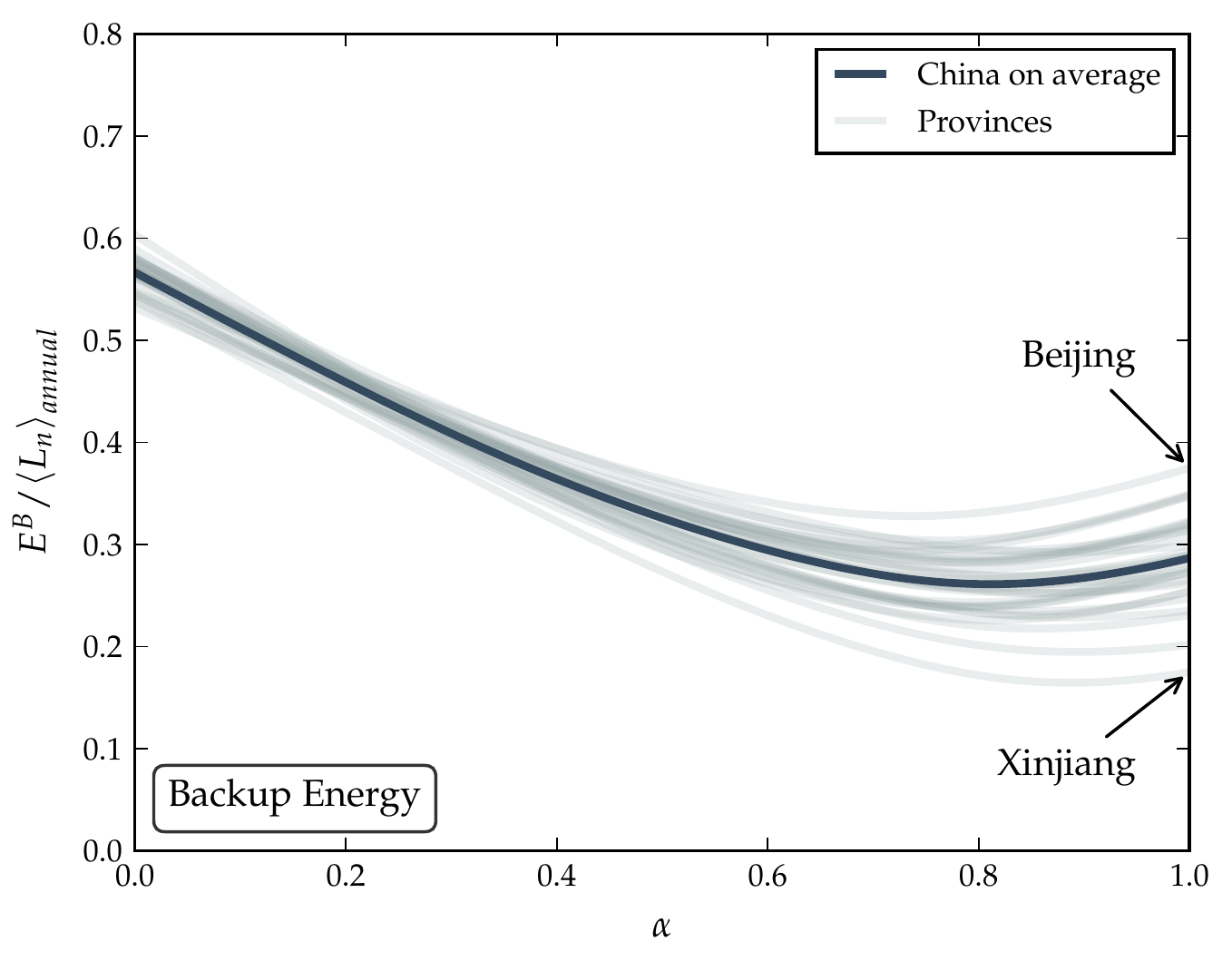}
	\caption{Backup Energy $E^B $ in units of average annual load as a function of the wind-solar mix $\alpha$. The fully drawn line represents the result for China on average with no transmission, while the faded lines indicate the results for each province.}
	\label{fig:homo_backup_energy}
\end{figure}

\begin{figure}[h!]
	\centering
	\includegraphics[width=0.6\linewidth]{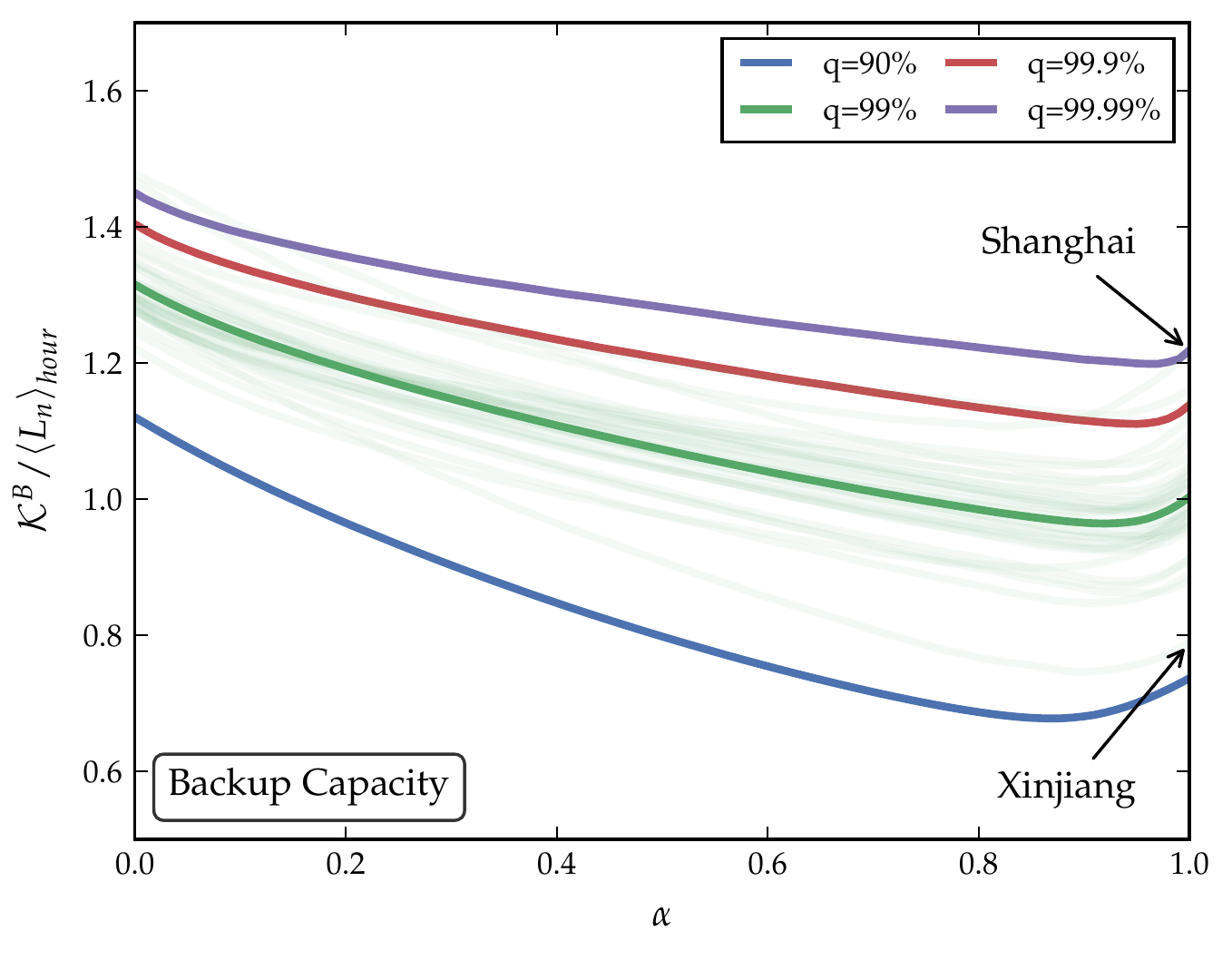}
	\caption{Backup Capacity $ \mathcal{K}^B $ in units of average hourly load as a function of the wind-solar mix $\alpha$. Four different quantiles of the backup time series are used to give estimates of $ \mathcal{K}^B $. The fully drawn lines indicate the result for averaged Chinese provinces with no transmission, while the faded lines represent the results of each province for $ q=0.99 $ only.}
	\label{fig:homo_backup_cap}
\end{figure}

\begin{figure*}[h!]
	\centering
	\includegraphics[width=\linewidth]{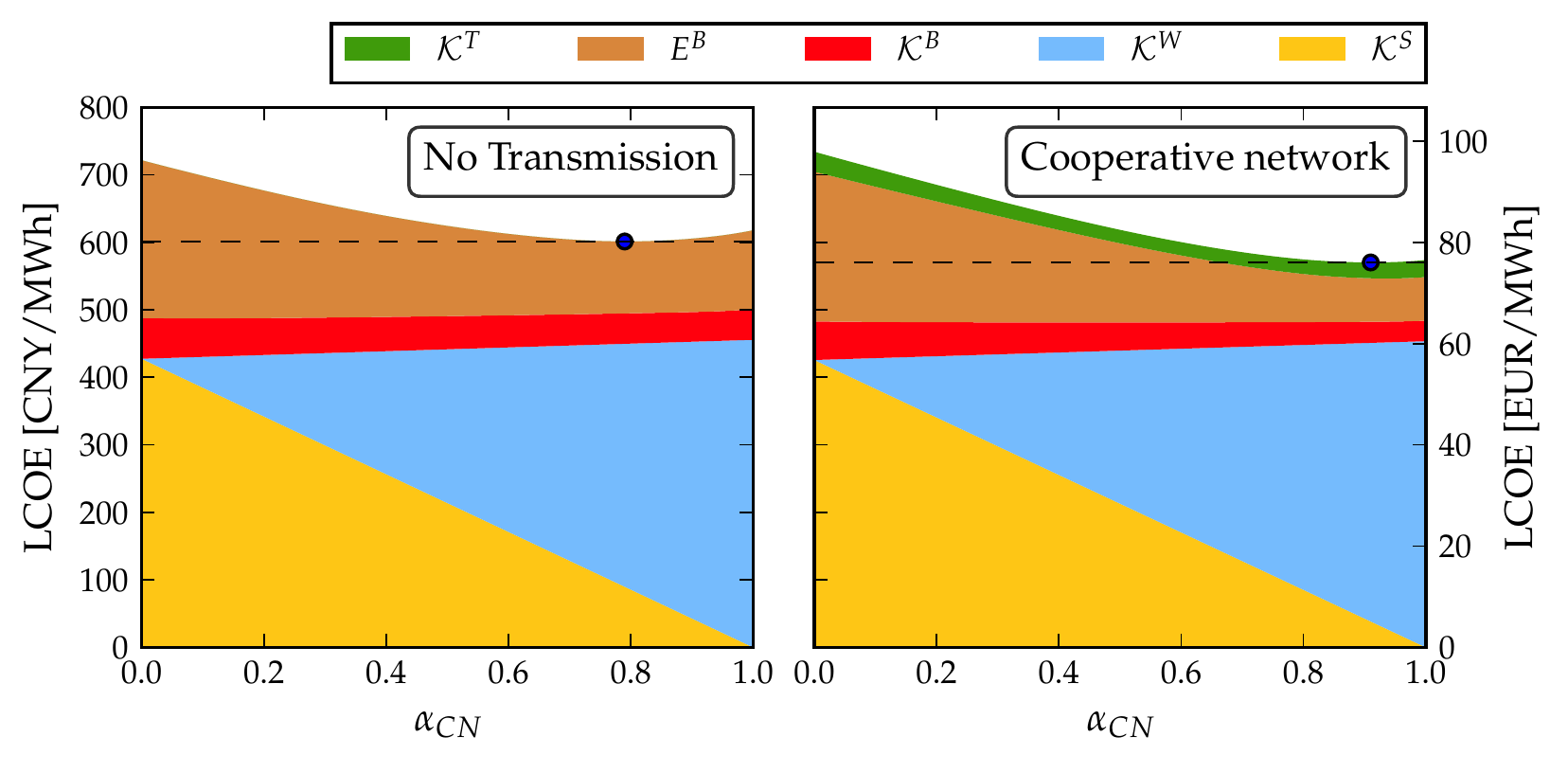}
	\caption{Component-wise Levelized Cost of Electricity (LCOE) as a function of the Chinese wind-solar mix $\alpha_{CN}$ in a zero-transmission (left) and cooperative homogeneous system (right). A bullet point and a dashed line denote the minimal cost in each case.}
	\label{fig:homo_cost}
\end{figure*}

\begin{figure}[h!]
	\centering
	\includegraphics[width=0.9\linewidth]{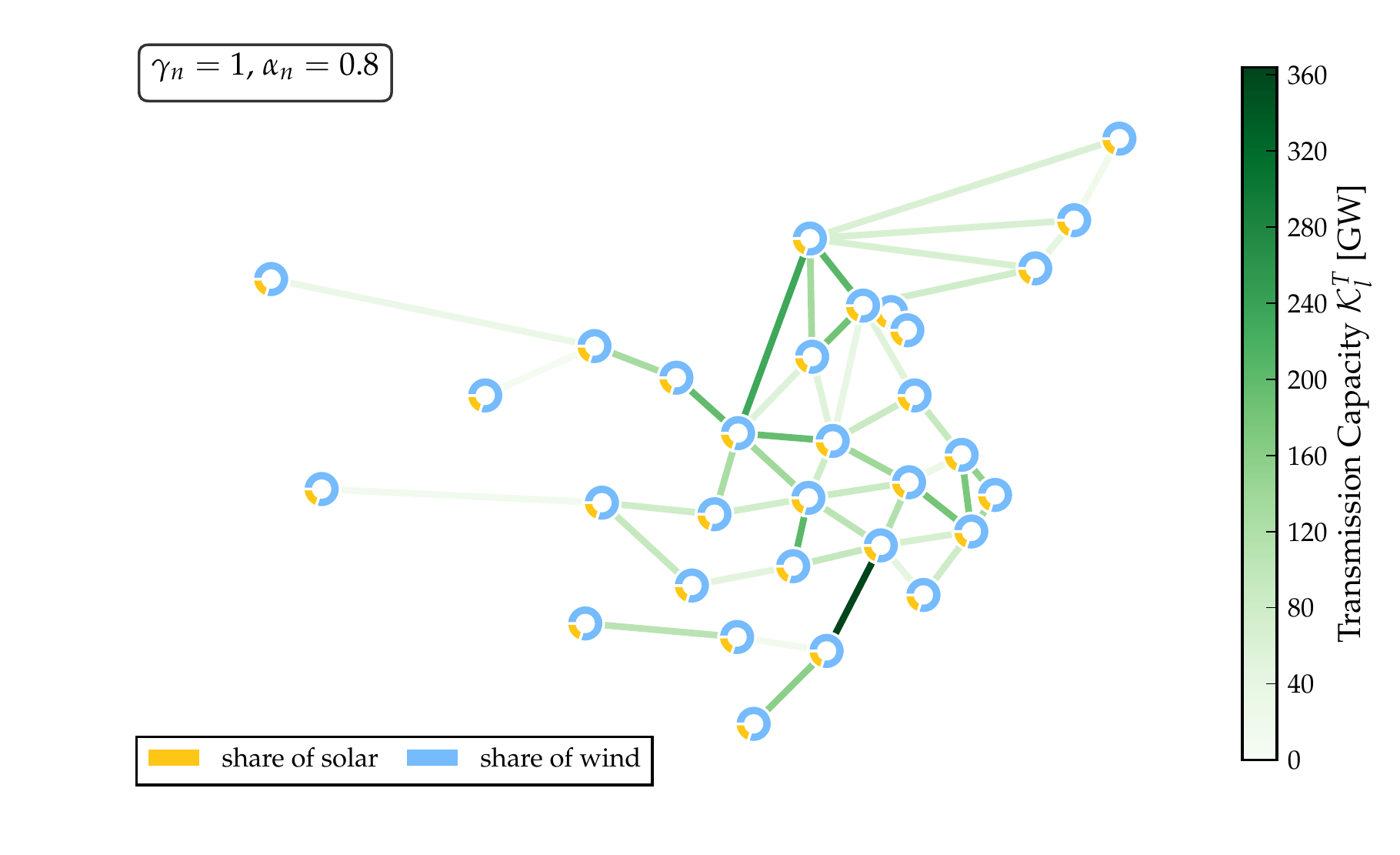}
	\caption{Transmission capacities $ \mathcal{K}^T_l$ in the homogeneous network. The nodal wind-solar production mix $ \alpha=0.8 $ is indicated by the colors blue (wind) and yellow (solar). The green lines correspond to the transmission capacity, scaled by the colorbar.}
	\label{fig:transusge_homo}
\end{figure}

Without transmission, by gradually increasing $\alpha $ from 0 to 1 in all isolated provinces, the share of wind power increases from 0 to 100\%. Under each scenario of $ \alpha $, the hourly mismatch between load and renewable generation is determined by Equation \ref{eq:mismatch}, and subsequently compensated by curtailment or backup energy in individual provinces. 

As shown in Figures \ref{fig:homo_backup_energy} and \ref{fig:homo_backup_cap}, the resulting overall infrastructure measures differ to some extent between provinces (shaded lines). Total backup energy decreases gradually to approximately 0.25 of the average annual load, at around $ \alpha_n = 0.8 $, before it slightly rises again when approaching the wind only limit. Looking at provincial results, high solar share scenarios require almost the same level of relative backup energy, but high wind shares give distinctly different levels of backup energy among provinces. This implies that, in China, intermittency in solar power generation is similar among the provinces, and wind power fluctuations are more dramatic in some than other. The variation in backup capacity for different mixing parameters is less pronounced. While higher quantiles naturally give a higher backup capacity, $ \mathcal{K}^B $ from the 99\% quantile decreases from 1.3, of average hourly load, to just below 1.0 with an increasing wind share. Overall, in this homogeneous case, the power system favors wind power compared to solar, and the mix that minimizes the backup infrastructure is around $\alpha = 0.8 $.

As for levelized costs, shown in Figure \ref{fig:homo_cost}a, the minimum is also located at $ \alpha \approx 0.8 $. Capital investments of wind and solar capacities generally contribute the most, and a solar only system has lower renewable generation capital cost compared to the wind only case. This is different from studies for Europe \cite{rodriguez2015cost}. The difference may be attributed to China's lower latitude, which results in increased solar radiation on average. Costs originating from backup energy follows the trend in Figure~\ref{fig:homo_cost}a, at the mix $ \alpha_{CN} = 0.8 $, LCOE reaches a minimum of 601.3 CNY/MWh (80.2 EUR/MWh).

When transmission between the provinces is introduced, the mismatch $ \Delta_n(t) $ between load and renewable power generation is first balanced by importing or exporting electricity in the power network, and then if more is needed, backup units will be dispatched. In Figure \ref{fig:homo_cost}b, the LCOE for these simulations is shown as a function of the wind-solar mix. Costs of wind and solar generators is identical to those of Figure~\ref{fig:homo_cost}a, but the backup costs are generally lower since surplus from one province can now replace backup generation in another instead of being curtailed. A cost for the transmission network is naturally added. The total cost is different from Figure~\ref{fig:homo_cost}a. For high solar shares the cost is slightly higher while it is lower for the wind rich mixes, where the optimum cost is located. This difference between the solar and wind rich mixes is explained by the dominating spatial and temporal patterns of the resources. The minimum LCOE for the case with transmission is  574.8 CNY/MWh (76.6 EUR/MWh), which is about 5\% lower than that with no transmission. The optimal mix moves further towards a 100\% wind scenario, but since the minimum is flat  a similar low LCOE can also be obtained for $ \alpha_{CN} = 0.8 $.

Figure \ref{fig:transusge_homo} shows the resulting normalized transmission capacities of the lines for the homogeneous network with $ \alpha_n = \alpha = 0.8 \;\forall n $. The transmission capacities are not homogeneous across the network. The strongest links are the ones connecting the central region to provinces where large amount of excess renewable energy is produced, for example, Inner Mongolia, Guangdong and Zhejiang. These provinces can export so much electricity because they have large nodal loads and subsequently, on average, large wind and solar generation.

\section{Optimized design: the heterogeneous network}
\label{hetero}

The heterogeneity in resource quality as well as temporal weather patterns can be exploited to configure a better geographical distribution of the wind and solar power generation capacities \cite{thesis:leon}. This section aims to lower the levelized cost of electricity in the system further by varying the mixing and penetration parameters $ \alpha_n $ and $ \gamma_n $ for the 31 provinces. 

\subsection{Production cost minimization}

\begin{figure}[h!]
	\centering
	\includegraphics[width=\linewidth]{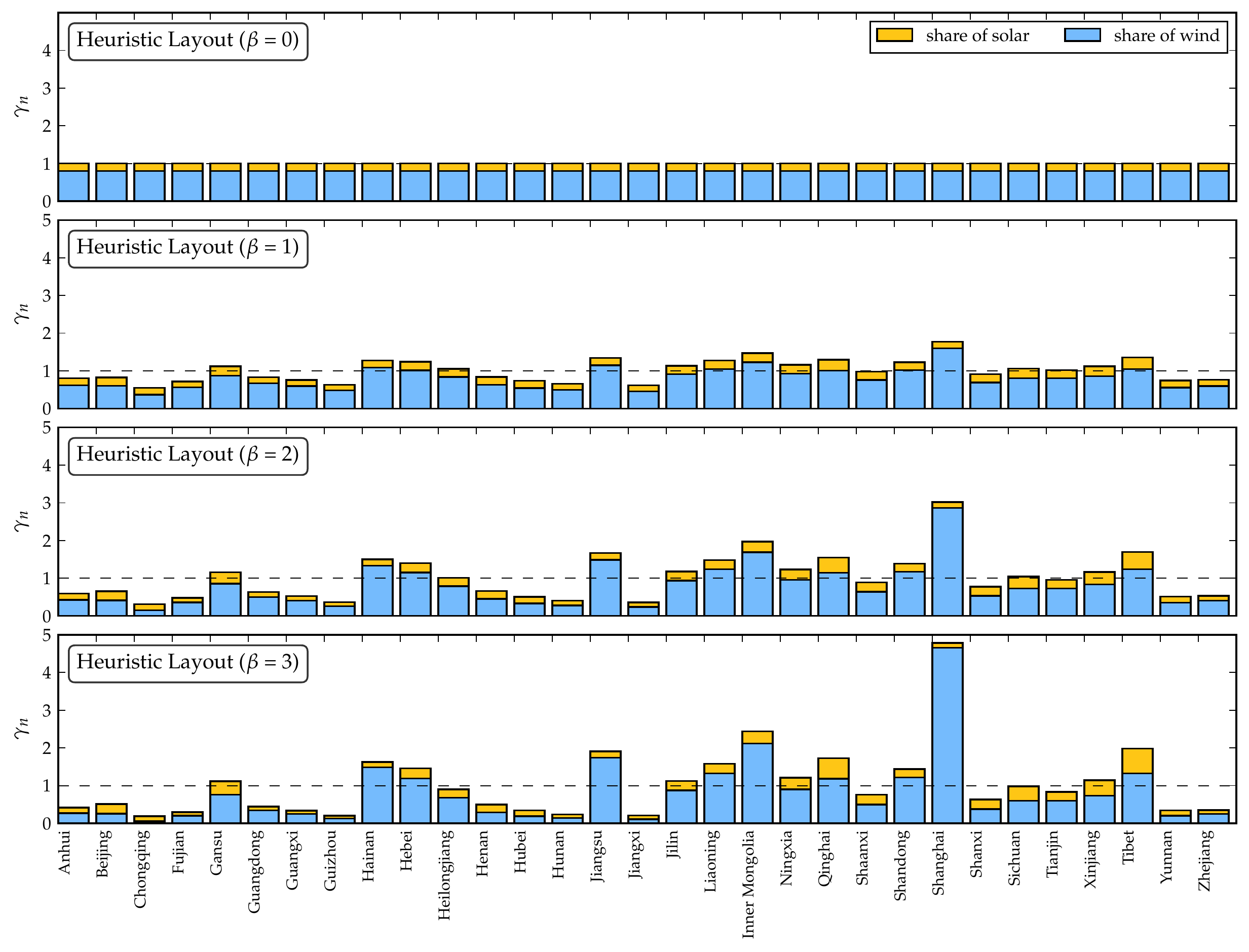}
	\caption{Heuristic $ \beta$-layouts where blue/yellow denote the wind/solar mix and bar heights represent the renewable penetration $ \gamma_n $. With $ \beta $ going from 0 to 3, the penetration levels gradually deviate from one, and the initial 80/20 mix also becomes heterogeneously distributed among provinces.}
	\label{fig:heuristicBetaLayout}
\end{figure}

Since the majority of the system cost in the baseline design derives from the wind and solar generators, the obvious way of lowering the total cost is by relocating them from low to high $ CF $ provinces. This may very well incur greater transmission costs and possibly also change the balancing costs due to different renewable production patterns. In this subsection, a heuristic that gradually increases heterogeneity by relocating wind and solar generators is introduced and the resulting system LCOE is explored. 

Using the optimal wind-solar mix $ \alpha_{CN} = 0.8 $  from the homogeneous layout, we assign arbitrary values 0, 1, 2, 3 to $ \beta $  in \Cref{eq:betalayout1,eq:betalayout2,eq:betalayout3,eq:betalayout4}. $ \beta=0 $ is the homogeneous baseline design we discussed in Section \ref{homo}, while $ \beta=1, 2, 3 $ allow provinces with higher $ CF $ bigger penetration levels than 100\% and others smaller. We call the these layouts heterogeneous as opposed to the baseline design. In this sense, $ \beta $ is the heterogeneity parameter. The results are shown in Figure~\ref{fig:heuristicBetaLayout}. $ \beta=0 $ is identical to the homogeneous baseline design discussed in Section \ref{homo}, while $ \beta=1, 2 $ or $ 3 $ allow provinces with high $ CF $ to increase their renewable penetration levels beyond 100\%, and provinces with relatively low $CF$ to make reductions.

Provinces with a high wind capacity factor $ CF^W $  like Shanghai, Inner Mongolia, Jiangsu and Hainan are assigned penetration levels well exceeding 100\%, while above average solar power penetrations are seen in provinces like Tibet, Qinghai and Xinjiang where solar radiation is stronger than average, as shown in Figure \ref{fig:CF_on_map}. 

This heuristic way of determining $ \gamma_n $ and $ \alpha_n $ produces a higher average capacity factor for China as a whole, and consequently lowers the total capital cost for renewable capacities. However, in addition to capital investments, LCOE also consists of costs from backup energy, backup infrastructure and transmission lines. The component-wise LCOE of the layouts is illustrated in the left part of Figure \ref{fig:opt_cost} and summarized in the supplementary material. As $ \beta $ changes from $ \beta = 0 $ to $ 3 $, the overall $  CF $ rises. Therefore the total capital investments for wind and solar power, shown in blue and yellow respectively, become smaller. At the same time, we also see a  more than 50\% decrease of backup energy, which can be credited to the smoothing effect caused by smaller correlations among high generation provinces, having a relatively large distance among them \cite{martin2015variability}. On the other hand, higher transmission costs emerge with larger $ \beta $ values. Indeed, most of the provinces assigned with high penetration levels are not close to high-demand regions, where wind/solar resources happen to be rather poor. This would require large-volume and long-distance transmission lines. In total, the LCOE monotonically decreases with $ \beta $.  As a fairly extreme scenario, $ \beta = 3 $ gives a LCOE of 477.6 CNY/MWh (63.7 EUR/MWh). This is almost a 17\% drop from the $ \beta = 0 $ layout.

\subsection{System cost minimization}

\begin{figure*}[h!]
	\centering
	\includegraphics[width=\linewidth]{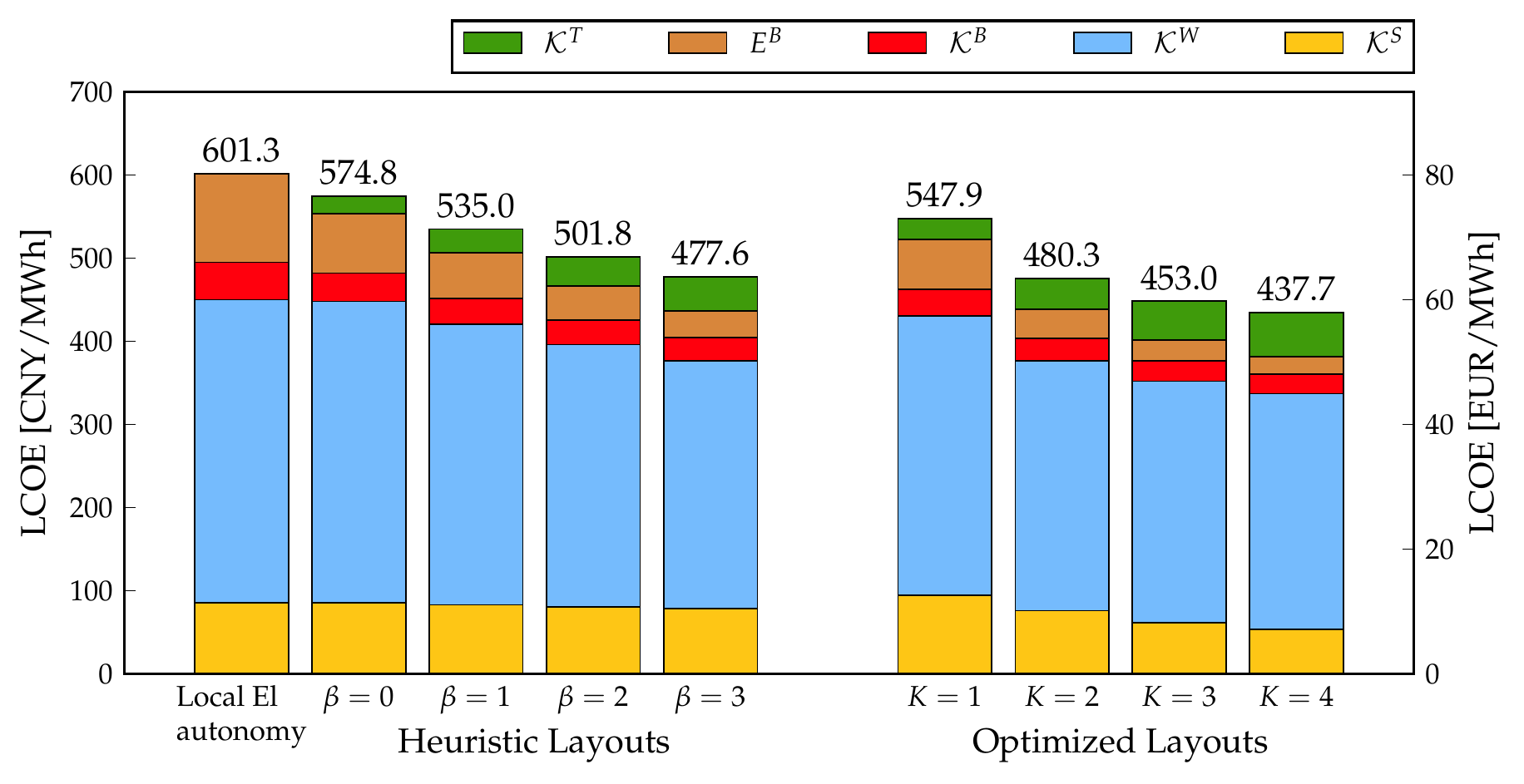}
	\caption{Component-wise LCOE for heuristic $ \beta $ and optimized layouts, as well as the  homogeneous layout with zero-transmission for comparison. Transmission costs are denoted by green bars, backup energy orange, backup capacity red, wind capacity blue and solar yellow. On top of the stacked bars the total LCOE is indicated in CNY/MWh.}
	\label{fig:opt_cost}
\end{figure*}

\begin{figure}[h!]
	\centering
	\includegraphics[width=\linewidth]{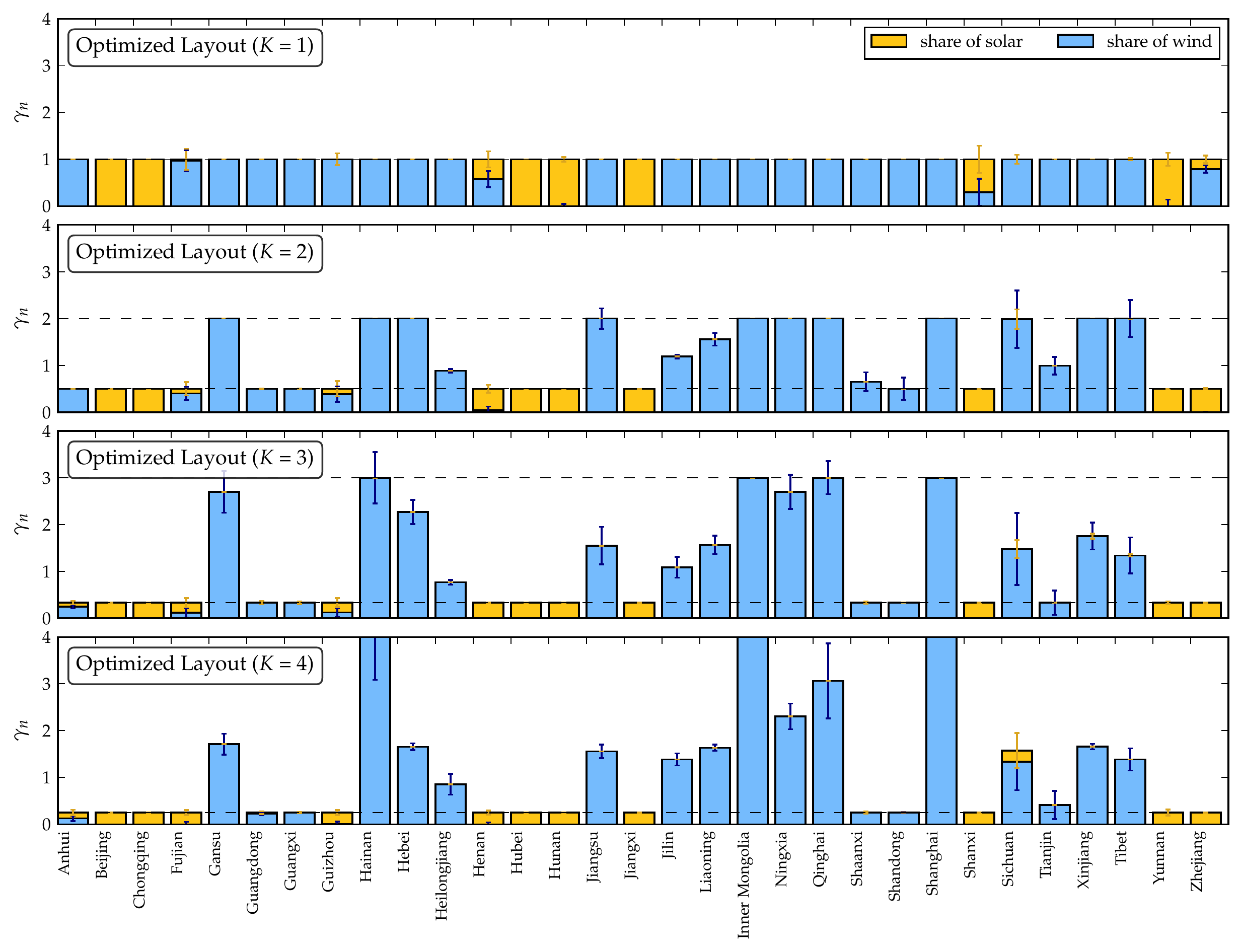}
	\caption{Optimized layouts with $ \frac{1}{K} $ / $ K $ as lower/upper bounds, where blue/yellow denote the wind/solar mix and bar heights represent the renewable penetration $ \gamma_n $. The error bars indicate the impact of 2005-2012 inter-year weather variability on optimal layouts.}
	\label{fig:GASLayout}
\end{figure}

The optimized layouts based on Greedy Axial Search, in Figure \ref{fig:GASLayout}, look quite different from the heuristic $ \beta $-layouts. Most provinces end up with wind or solar power only. This is because in most provinces wind and solar capacity factors are quite different. The  provinces of the central Southern region, for example, Chongqing, Henan, Hubei and Hunan, are characterized by very low $ CF^W $ but fairly good $ CF^S $, so the optimizer decides to install only solar panels there. As for wind-abundant provinces, like Hainan, Hebei, Jiangsu, Inner Mongolia and Xinjiang, the optimization successfully identifies them and suggests installing plenty of wind turbines well exceeding their own needs. 

Layout with $ K = 1 $ fixes all penetrations to $ \gamma_n=1 $, but wind-solar mixing is different among the provinces. In this sense it can be most directly compared to the $\beta$-layout with $\beta=0$. Relaxing $ \alpha_n $ in this way alone can drop the LCOE from 574.8 to 547.9 CNY/MWh (76.6 to 73.1 EUR/MWh) by 4.7\%. An obvious difference is in the overall share of solar power: the optimization also identifies the CF differences between solar and wind in the central region and decides to invest in solar power. Seen in Figure \ref{fig:opt_cost}a, this relaxation does not have a large impact on backup or transmission costs.

The optimized layouts with $ K = 2, 3$ or $4 $  manage to reduce the infrastructure costs further, seen in the right part of Figure \ref{fig:opt_cost}.  The costs for total backup energy $ E^B $  drops by up to 64\%, while backup capacity $ \mathcal{K}^B $ only decreases by a quarter. In contrast, transmission almost doubles from $ K=1 $ to $ K=4 $. This is attributed to the imports of the resource-poor provinces in the central region.

\subsection{Transmission topology}

\begin{figure*}[h!]
	\centering
	\includegraphics[width=1.1\linewidth]{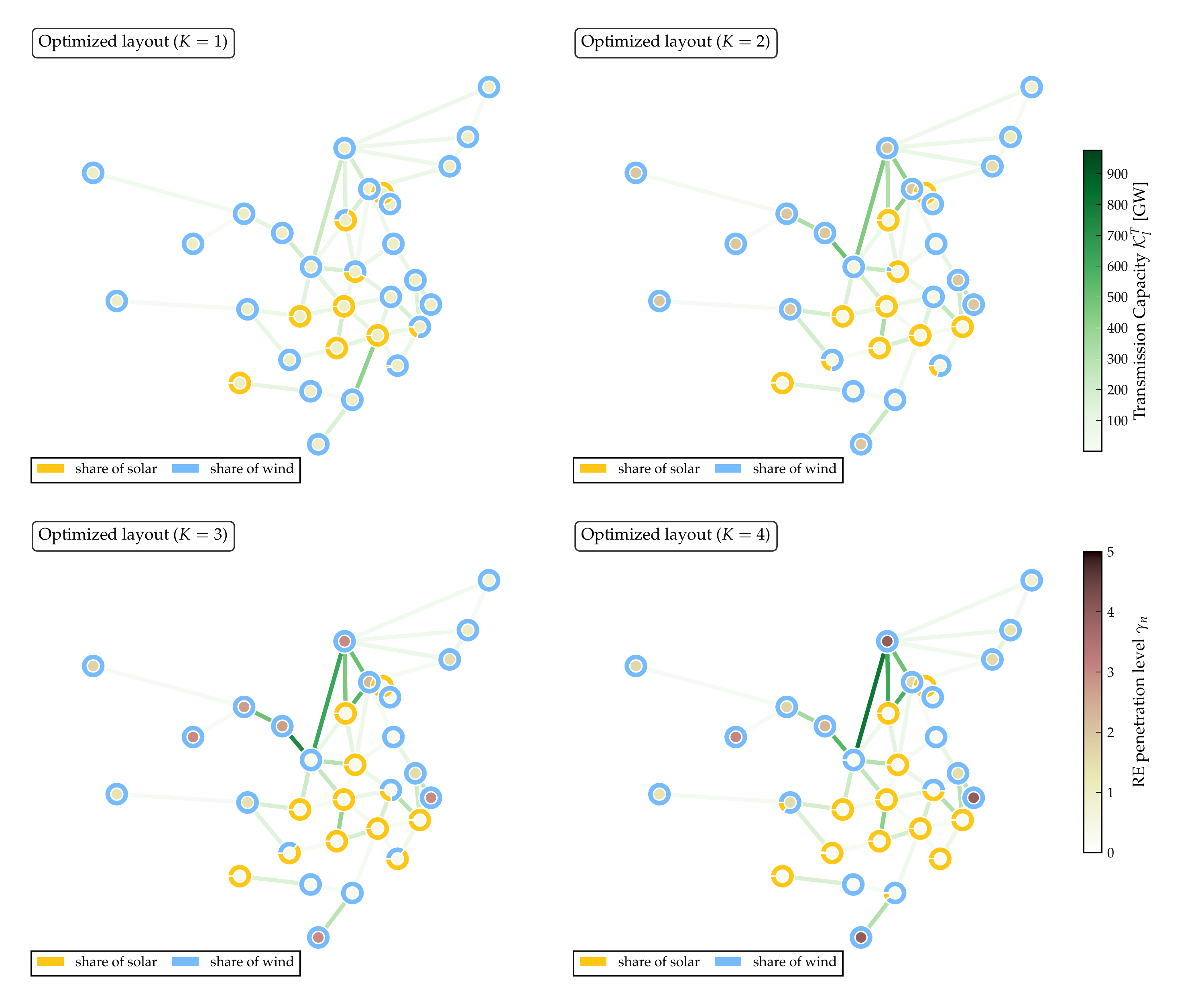}
	\caption{Transmission capacities $ \mathcal{K}^T_l $ in the heterogeneous optimized layouts. Renewable penetrations are represented by the colored discs in the nodes, and the green lines correspond to transmission volume. The rings with blue/yellow color show the average nodal production mix of wind and solar.}
	\label{fig:TransmissionUsage_gas_KT}
\end{figure*}

Taking a closer look at transmission usage in the optimized electricity network, Figure~\ref{fig:TransmissionUsage_gas_KT} visualizes the geographical distribution of  transmission capacity $ \mathcal{K}^T_l $ on the links, mixing parameters $ \alpha_n $, as well as renewable penetration $ \gamma_n $. It is evident that a larger $ K $ increases the spatial heterogeneity in transmission capacity layouts. $ K=1 $ gives a fairly homogeneous transmission system, especially in the central Eastern regions, with the exception of the line connecting the high-load province Guangdong with the central provinces. This is  because its southern neighbors do not have high power demands, and this is the main path to export its excess electricity.

As $ K $ grows larger, transmission demand starts to shift towards the lines connecting the northwestern provinces and the central region. This can be attributed to the high renewable penetration levels in these provinces, although all of them have fairly low average load except Inner Mongolia. 

It seems after increasing heterogeneity, Inner Mongolia becomes the dominating source of renewable power in China. With high penetration levels and its second largest average load, it virtually becomes the powerhouse of the country. Actually in recent years, Inner Mongolia has had the largest wind power capacity. By the end of 2015, for example, 17.6\% of China's wind capacity had been installed in Inner Mongolia  \cite{wind2015}.

As to the East coast, Jiangsu and Shanghai are assigned fairly high penetration levels because of their potentially high offshore or near-shore wind capacities. Benefiting from this, their neighboring provinces, Zhejiang, Anhui and Shandong, do not need much renewable capacities despite high annual loads. In turn, they can rely on imports, which can be corroborated by the heavy transmission usage among them.

\subsection{Sensitivity analysis}

\begin{figure}[h!]
	\centering
	\includegraphics[width=\linewidth]{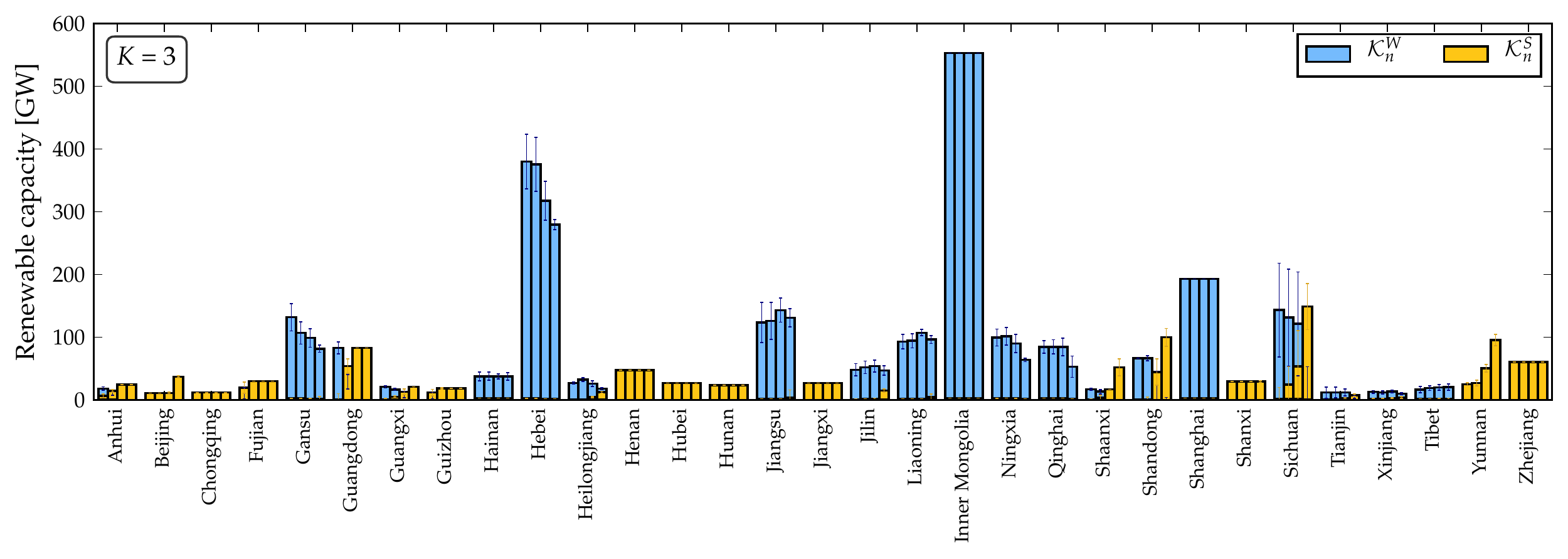}
	\caption{Optimized renewable capacity layouts where blue/yellow denote the wind/solar mix and bar heights represent the renewable capacity $ \mathcal{K}_n^{W+S} $. For each province, the four bars indicate the scenarios for solar cost reduction of 0\%, 10\%, 25\%, 50\%, respectively. The error bars indicate the impact of 2005-2012 inter-year weather variability on optimal layouts.}
	\label{fig:GASLayout_sensitivity_K3}
\end{figure}

While onshore wind power cost has remained relatively stable in recent years, PV module predominated solar power cost has dropped by 75\% from 2008 \cite{mueller2016next}. This can largely be attributed to China's massive growth in solar panel production and installation. Along with both state and provincial level subsidies for initial investments, centralized and distributed PV, installed solar capacities have shown significant increases in the past three years \cite{yearbook}.
Considering the colossal market and sustained policy incentives, cost of solar PV may see an further reduction until 2050 \cite{mueller2016next}. Here, to analyze the sensitivity to future price drops, we calculated the optimized layouts for solar cost reductions of 10\%, 25\% and 50\%. For the $ K=3 $ layout, absolute renewable capacity under the four price drop scenarios is shown in Figure \ref{fig:GASLayout_sensitivity_K3}. The trend of wind-dominated generation is not changed despite up to halved solar cost, especially in provinces with high assigned penetrations. However, in Anhui, Sichuan, Guangdong, Shandong and Shaanxi, the shift to a gradually increasing solar capacity is clear. Their relative low wind capacity factors gave cheap local PV the advantage, even with increased backup and transmission cost. 

Inter-year weather pattern variability can also have an impact on the renewable capacity distribution in the optimized layouts, and its sensitivity is indicated as error bars in Figures \ref{fig:GASLayout} and \ref{fig:GASLayout_sensitivity_K3}.

\section{Discussion}
\label{disc}

Both the heuristic and optimized layouts show that shifting more renewable capacity to the Northwest would result in a sizable reduction in total system cost, provided a well-connected transmission grid is also installed. This is in good agreement with the government's current renewable policy and long term energy strategy \cite{hua2016development}. A national shift to wind and solar power is an opportunity, in our opinion, for the provinces with high renewable potential to attract external investments and high-tech human resources. A well-developed top-down, production-installation-maintenance renewable industrial chain would play a major role in the region's economic prosperity.

Today, curtailment is what's burdening the region's renewable development. With the vast majority of the country's wind turbines and solar PV located in Inner Mongolia, Gansu and Xinjiang, renewable development has been burdened by up to 30\% curtailment rate, even with a mere 5\% overall penetration \cite{he2017china}. This has mainly been attributed to regional market barriers and poor transmission planning \cite{luo2016wind}. In Inner Mongolia, for example, wind power can meet almost 100\% of the local electricity demand during the day, but the local CHPs are given priority to run at night instead of wind farms to provide district heating to residents, in spite of the usually stable and excellent wind resources over the plateau. Realizing this, the National Development and Reform Commission and the National Energy Administration have taken major actions such as ceasing approval of new wind farm or centralized PV stations in Northwestern provinces until curtailment decrease, breaking up the five regional TSOs for a better control of transmission planning and construction, subsidizing distributed solar PV and so on.

In the far future renewable power system, described here, curtailment or alternative use of the surplus energy is unavoidable. With an overall penetration of 100\%, since the average renewable generation equals the average load, the total curtailed energy is equal to the total backup energy. Consulting Figure \ref{fig:opt_cost}, overall curtailment goes down with higher heterogeneity parameters $ \beta, K $, following the same trend as LCOE. For better resource utilization, though, curtailed electricity can otherwise be stored or sold to transportation and heating sectors. In this regard, Li et al. \cite{li2016electric} found that controlled charging of electric vehicles has significant advantages in mitigating grid mismatch and facilitating renewable generation in China. Zhang et al. \cite{zhang2016reducing} identified electric boilers as a cost-effective medium to transform excess renewable power to heat in a unit commitment dispatch model. Their suitability for large-scale expansion also makes these possibilities very promising.

\section{Conclusion and outlook}
\label{conc}

In this paper, a future Chinese electricity system with an average wind and solar power generation equal to the national load, i.e., 100\% renewable penetration, is analyzed in terms of the provincial wind-solar production mix and penetration, the required inter-provincial transmission network, dispatchable backup energy and capacity requirements and total system cost. A high spatio-temporal resolution dataset for wind and solar generation as well as hourly provincial demand is developed to enable the study. 

By locating all wind and solar capacities directly proportional to the annual demand in the individual provinces and disallowing transmission between provinces, we find a mix of 80\% wind and 20\% solar to be optimal in terms of backup energy and levelized cost. On the other hand, a power network interconnecting provincial nodes, reduces total backup needs and lowers the system LCOE by 5\% despite additional transmission investments for this so-called homogeneous baseline.

The majority of the system cost comes from wind turbines and solar PVs, while their capacity installation depends on local capacity factors. Allowing the system to build more renewables in resource-rich regions means that fewer wind turbines and solar panels are needed in order to produce the same amount of energy. This leads to heterogeneous layouts where provincial penetration $ \gamma_n $ varies. The two methods, i.e. production cost minimization and system cost minimization, distribute wind and solar capacities heterogeneously over the whole country, and they manage to lower not only total capacity cost but also backup generation in the power network. By raising heterogeneity to $ K=4 $, system LCOE is reduced by 27\%, compared to the baseline scenario, i.e., down to 437.7 CNY/MWh (58.4 EUR/MWh).

A heterogeneous system also changes the topology of the transmission network. Higher capacities are needed in links connecting provinces like Inner Mongolia, Xinjiang in the Northwest and the central region. The former, resource-rich region supplies its excess clean energy to the central high-load provinces. This peculiar demand/resource distribution could be an interesting topic to explore further, given the economic disparity between them.

Currently the renewable sector in China is burdened by high curtailment, despite a relativly low penetration of about 5\%. In a 100\% renewable system surplus electricity is unavoidable, which means that the energy can be either curtailed or used in alternative ways. To fully exploit the renewable potential, the energy could be used by coupling the power sector to, e.g., district heating or transportation considering the growing popularity of electric vehicles. 

Furthermore, integrating hydroelectricity into the wind-solar only system may further reduce conventional backup as well as transmission volume, since hydro dams are clustered in the southeastern provinces close to high demand regions.

\hspace{0.5cm}

\section*{Acknowledgments}

The first author gratefully acknowledges the financial support from the Chinese Scholarship Council. Gorm B. Andresen and Martin Greiner are partially funded by the RE-INVEST project (Renewable Energy Investment Strategies -- A two-dimensional interconnectivity approach), which is supported by Innovation Fund Denmark (6154-00022B). The responsibility for the contents lies solely with the authors.

\section*{Bibliography}
\bibliographystyle{unsrt}
\bibliography{revision_for_Energy.bib}

\begin{thebibliography}{10}

\bibitem{huang2012typical}
Kan Huang, Guoshun Zhuang, Yanfen Lin, JS~Fu, Q~Wang, T~Liu, R~Zhang, Y~Jiang,
  C~Deng, Q~Fu, et~al.
\newblock {Typical types and formation mechanisms of haze in an Eastern Asia
  megacity, Shanghai}.
\newblock {\em Atmospheric Chemistry and Physics}, 12(1):105, 2012.

\bibitem{yearbook}
CEP.
\newblock {\em {China Electric Power Yearbook}}.
\newblock China Electric Power Press, 2006-2015.

\bibitem{Analysis}
{State Grid Energy Research Institute}.
\newblock {\em {Analysis report on renewable energy generation in China 2016}}.
\newblock China Electric Power Press, 2016.

\bibitem{brown}
Tom Brown, Stefan Langanke, Thomas Ackemann, and Sven Teske.
\newblock {Integrating Renewables in Jiangsu Province, China}.
\newblock In {\em {The 14th International Workshop on Integration of Wind Power
  into Power Systems}}, 2015.

\bibitem{he2014and}
Gang He and Daniel~M Kammen.
\newblock {Where, when and how much wind is available? A provincial-scale wind
  resource assessment for China}.
\newblock {\em Energy Policy}, 74:116--122, 2014.

\bibitem{he2016and}
Gang He and Daniel~M Kammen.
\newblock {Where, when and how much solar is available? A provincial-scale
  solar resource assessment for China}.
\newblock {\em Renewable Energy}, 85:74--82, 2016.

\bibitem{zhang2017integrated}
Ning Zhang, Zhaoguang Hu, Bo~Shen, Gang He, and Yanan Zheng.
\newblock {An integrated source-grid-load planning model at the macro level:
  Case study for China's power sector}.
\newblock {\em Energy}, 126:231--246, 2017.

\bibitem{yi2016inter}
Bo-Wen Yi, Jin-Hua Xu, and Ying Fan.
\newblock {Inter-regional power grid planning up to 2030 in China considering
  renewable energy development and regional pollutant control: A multi-region
  bottom-up optimization model}.
\newblock {\em Applied Energy}, 184:641--658, 2016.

\bibitem{huber2015optimal}
Matthias Huber and Christoph Weissbart.
\newblock {On the optimal mix of wind and solar generation in the future
  Chinese power system}.
\newblock {\em Energy}, 90:235--243, 2015.

\bibitem{rodriguez2015cost}
Rolando~A Rodriguez, Sarah Becker, and Martin Greiner.
\newblock {Cost-optimal design of a simplified, highly renewable pan-European
  electricity system}.
\newblock {\em Energy}, 83:658--668, 2015.

\bibitem{schlachtberger2017benefits}
D.P. Schlachtberger, T.~Brown, S.~Schramm, and M.~Greiner.
\newblock {The benefits of cooperation in a highly renewable European
  electricity network}.
\newblock {\em Energy}, 134:469--481, 2017.

\bibitem{becker2014features}
Sarah Becker, Bethany~A Frew, Gorm~B Andresen, Timo Zeyer, Stefan Schramm,
  Martin Greiner, and Mark~Z Jacobson.
\newblock {Features of a fully renewable US electricity system: Optimized mixes
  of wind and solar PV and transmission grid extensions}.
\newblock {\em Energy}, 72:443--458, 2014.

\bibitem{becker2015renewable}
Sarah Becker, Bethany~A Frew, Gorm~B Andresen, Mark~Z Jacobson, Stefan Schramm,
  and Martin Greiner.
\newblock {Renewable build-up pathways for the US: Generation costs are not
  system costs}.
\newblock {\em Energy}, 81:437--445, 2015.

\bibitem{prasad2017assessment}
Abhnil~A Prasad, Robert~A Taylor, and Merlinde Kay.
\newblock {Assessment of solar and wind resource synergy in Australia}.
\newblock {\em Applied Energy}, 190:354--367, 2017.

\bibitem{andresen2015validation}
Gorm~B Andresen, Anders~A S{\o}ndergaard, and Martin Greiner.
\newblock {Validation of Danish wind time series from a new global renewable
  energy atlas for energy system analysis}.
\newblock {\em Energy}, 93:1074--1088, 2015.

\bibitem{staffell2016using}
Iain Staffell and Stefan Pfenninger.
\newblock {Using bias-corrected reanalysis to simulate current and future wind
  power output}.
\newblock {\em Energy}, 114:1224--1239, 2016.

\bibitem{pfenninger2016long}
Stefan Pfenninger and Iain Staffell.
\newblock {Long-term patterns of European PV output using 30 years of validated
  hourly reanalysis and satellite data}.
\newblock {\em Energy}, 114:1251--1265, 2016.

\bibitem{wiki:network}
Wikipedia.
\newblock {Ultra high voltage electricity transmission in China}, 2017.
\newblock [Online; accessed 21-April-2017].

\bibitem{thesis:leon}
Emil~H. Eriksen, Leon~J. Schwenk-Nebbe, Bo~Tranberg, Tom Brown, and Martin
  Greiner.
\newblock {Optimal heterogeneity in a simplified highly renewable European
  electricity system}.
\newblock {\em Energy}, 133:913--928, 2017.

\bibitem{wind2015}
{Chinese Wind Energy Association}.
\newblock {Summary of installed wind capacity in China 2015}.
\newblock {\em Wind Energy (In Chinese)}, 2:48--63, 2016.

\bibitem{rodriguez2014transmission}
Rolando~A Rodriguez, Sarah Becker, Gorm~B Andresen, Dominik Heide, and Martin
  Greiner.
\newblock {Transmission needs across a fully renewable European power system}.
\newblock {\em Renewable Energy}, 63:467--476, 2014.

\bibitem{lin2016economic}
Jiang Lin, Gang He, and Alexandria Yuan.
\newblock {Economic rebalancing and electricity demand in China}.
\newblock {\em The Electricity Journal}, 29(3):48--54, 2016.

\bibitem{Domestic82:online}
OECD.
\newblock {Domestic product - GDP long-term forecast - OECD Data}.
\newblock \url{https://data.oecd.org/gdp/gdp-long-term-forecast.htm}.
\newblock (Accessed on 12/24/2016).

\bibitem{rodriguez2015localized}
Rolando~A Rodriguez, Magnus Dahl, Sarah Becker, and Martin Greiner.
\newblock {Localized vs. synchronized exports across a highly renewable
  pan-European transmission network}.
\newblock {\em Energy, Sustainability and Society}, 5(1):21, 2015.

\bibitem{tranberg2015power}
Bo~Tranberg, Anders~B Thomsen, Rolando~A Rodriguez, Gorm~B Andresen, Mirko
  Sch{\"a}fer, and Martin Greiner.
\newblock {Power flow tracing in a simplified highly renewable European
  electricity network}.
\newblock {\em New Journal of Physics}, 17(10):105002, 2015.

\bibitem{ouyang2014impacts}
Xiaoling Ouyang and Boqiang Lin.
\newblock {Impacts of increasing renewable energy subsidies and phasing out
  fossil fuel subsidies in China}.
\newblock {\em Renewable and Sustainable Energy Reviews}, 37:933--942, 2014.

\bibitem{ouyang2014levelized}
Xiaoling Ouyang and Boqiang Lin.
\newblock {Levelized cost of electricity (LCOE) of renewable energies and
  required subsidies in China}.
\newblock {\em Energy policy}, 70:64--73, 2014.

\bibitem{wangweb}
Zhonghong Wang.
\newblock {HVDC and AC}, 2014.
\newblock [Accessed: 2017-04-18].

\bibitem{martin2015variability}
Clara M~St Martin, Julie~K Lundquist, and Mark~A Handschy.
\newblock Variability of interconnected wind plants: correlation length and its
  dependence on variability time scale.
\newblock {\em Environmental Research Letters}, 10(4):044004, 2015.

\bibitem{mueller2016next}
S~Mueller, P~Frankl, and K~Sadamori.
\newblock Next generation wind and solar power from cost to value.
\newblock {\em International Energy Agency: Paris, France}, 2016.

\bibitem{hua2016development}
Yaping Hua, Monica Oliphant, and Eric~Jing Hu.
\newblock {Development of renewable energy in Australia and China: A comparison
  of policies and status}.
\newblock {\em Renewable Energy}, 85:1044--1051, 2016.

\bibitem{dong2016spatial}
Liang Dong, Hanwei Liang, Zhiqiu Gao, Xiao Luo, and Jingzheng Ren.
\newblock {Spatial distribution of China's renewable energy industry: Regional
  features and implications for a harmonious development future}.
\newblock {\em Renewable and Sustainable Energy Reviews}, 58:1521--1531, 2016.

\bibitem{he2017china}
Gang He, Hongliang Zhang, Yuan Xu, and Xi~Lu.
\newblock China’s clean power transition: current status and future prospect.
\newblock {\em Resources, Conservation and Recycling}, 121:3--10, 2017.

\bibitem{luo2016wind}
Guoliang Luo, Yanling Li, Wenjun Tang, and Xiao Wei.
\newblock {Wind curtailment of China's wind power operation: Evolution, causes
  and solutions}.
\newblock {\em Renewable and Sustainable Energy Reviews}, 53:1190--1201, 2016.

\bibitem{li2016electric}
Ying Li, Chris Davis, Zofia Lukszo, and Margot Weijnen.
\newblock {Electric vehicle charging in China’s power system: Energy,
  economic and environmental trade-offs and policy implications}.
\newblock {\em Applied energy}, 173:535--554, 2016.

\bibitem{zhang2016reducing}
Ning Zhang, Xi~Lu, Michael~B McElroy, Chris~P Nielsen, Xinyu Chen, Yu~Deng, and
  Chongqing Kang.
\newblock {Reducing curtailment of wind electricity in China by employing
  electric boilers for heat and pumped hydro for energy storage}.
\newblock {\em Applied Energy}, 184:987--994, 2016.

\end{thebibliography}

\end{document}